\documentclass[prd,twocolumn,lengthcheck,superscriptaddress,showpacs,letterpaper]{revtex4}

\usepackage{color}
\usepackage{graphicx}  
\usepackage{float}
\usepackage{amsmath}
\usepackage{subfigure}
\usepackage{hyperref}

\newcommand\HideSection[1]{}
\newcommand\HideSubSection[1]{}
\newcommand\optional[1]{}

\begin{document}

\title{A single-spin precessing gravitational wave in closed form}

\author{Andrew Lundgren}
\affiliation{Albert-Einstein-Institut, Callinstr. 38, 30167 Hannover, Germany}
\author{R. O'Shaughnessy}
\affiliation{Center for Gravitation and Cosmology, University of Wisconsin-Milwaukee, Milwaukee, WI 53211, USA}

\begin{abstract}
In coming  years, gravitational wave detectors should  find   black hole-neutron star
binaries, potentially coincident with astronomical phenomena like short GRBs.  These binaries are expected to precess.
Gravitational wave science requires a tractable model for precessing binaries, to disentangle precession physics from other
phenomena like modified strong field gravity,  tidal deformability, or Hubble flow;   and to measure compact object
masses, spins, and alignments.  
Moreover, current searches for gravitational waves from compact binaries use templates where the binary does not
precess and are ill-suited for detection of generic precessing sources.  
In this paper we provide a closed-form representation of the single-spin precessing waveform in the frequency domain by
reorganizing the signal as a sum over harmonics, each of which resembles a nonprecessing waveform.  
This form enables simple analytic calculations (e.g., a Fisher matrix) with easily-interpreted results.   
We have verified that for generic BH-NS binaries, our model agress with the time-domain waveform to 2\%.  
Straightforward extensions  of the derivations outlined here [and provided in full online] allow higher accuracy and error estimates.

\end{abstract}
\maketitle

\HideSection{Introduction}

For the astrophysically most plausible strong gravitational wave sources -- coalescing compact binaries of black holes and neutron stars  with total mass
$M=m_1+m_2\le 16 M_\odot$ --    ground based gravitational wave detector networks (notably LIGO \cite{gw-detectors-LIGO-original} and Virgo
\cite{Accadia:2011zz,aVIRGO})  are principally sensitive to each binary's nearly-adiabatic and quasi-circular inspiral
  \cite{2003PhRvD..67j4025B,2004PhRvD..70j4003B,2004PhRvD..70f4028D,BCV:PTF,2005PhRvD..72h4027B,2006PhRvD..73l4012K,2008PhRvD..78j4007H,gw-astro-mergers-approximations-SpinningPNHigherHarmonics,gw-astro-PN-Comparison-AlessandraSathya2009}.  
For nonprecessing binaries, the inspiral signal model has been explored in detail, both analytically and with numerical
simulation.  Fast, stationary-phase approximations to the signal exist which faithfully reproduce the signal \cite{TaylorT2F2Faithful}.
By contrast, plausible astrophysical processes will produce merging black hole binaries with arbitrary spin orientations, fully
populating the 15-dimensional model space for quasi-circular inspiral.   
Generic binaries orbits' both shrink and rapidly precess  \cite{ACST,2004PhRvD..70l4020S,gwastro-mergers-PNLock-Gerosa2013}.   
For many astrophysically interesting sources, including many binaries containing at least one black hole, significant spin precession is expected
while the signal passes through LIGO and Virgo's sensitive frequency band.  
Precession-induced modulations compete with effects from all other phenomena, including modifications to gravity or the
nature of nuclear matter.  Precession breaks severe degeneracies \cite{1995PhRvD..52..848P} and allows high-precision measurements of binary
parameters including high-precision mass, spin, and spin-orbit misalignment distributions, providing high-precision
constraints on astrophysical processes like the central engines of short GRBs.  
Precessing sources have been substantially less-thoroughly modeled.  At present, the current state-of-the-art
\texttt{SpinTaylorT4} \cite{BCV:PTF,gw-astro-PN-Comparison-AlessandraSathya2009} and \texttt{SpinTaylorT2} models
\cite{TaylorT2F2Faithful} solve  coupled time-domain ordinary
 differential equations to evolve the orbits and spins.    
Until this work, no fast,
stationary-phase approximation existed which faithfully reproduces the signal from a generic single-spin binary.  
A frequency-domain signal provides an analytically tractable tool for theoretical analysis; for example, the Fisher matrix  can  be efficiently evaluated and its physical significance understood
\cite{1995PhRvD..52..848P}.    
Also, these frequency-domain waveforms can be evaluated and explored with significantly reduced computational cost.  
Previous searches  balanced accuracy and coverage against computational cost, usually adopting low-cost but
easily-understood approximate
waveforms (``detection templates''), though others proposed the use of accurate but expensive time-domain signals
\cite{BuonannoChenVallisneri:2003a,2004PhRvD..70j4003B,BCV:PTF,2003PhRvD..67j4025B,2005PhRvD..72h4027B}.  
Present searches explicitly omit precession, hoping to be efficient at finding precessing binaries.  
Our   revised signal model may change the balance, allowing direct searches for precessing sources.  

In this paper, we show that the gravitational wave signal for a single-spin binary \footnote{We expect our scheme  approximates
  the evolution of two-spin systems with moderate mass ratio, since the second spin's influence on the orbit can be neglected.}
system can be well approximated with a simple analytic form in the frequency domain.  Our decomposition expresses the signal as a sum of five terms, each a weak modulation (sideband) of a
leading-order (``carrier'') function. 
Our decomposition relies on simple precession \cite{ACST} of the angular momenta and breaks down precisely when that
approximation does.  

\begin{figure}
\includegraphics[width=\columnwidth]{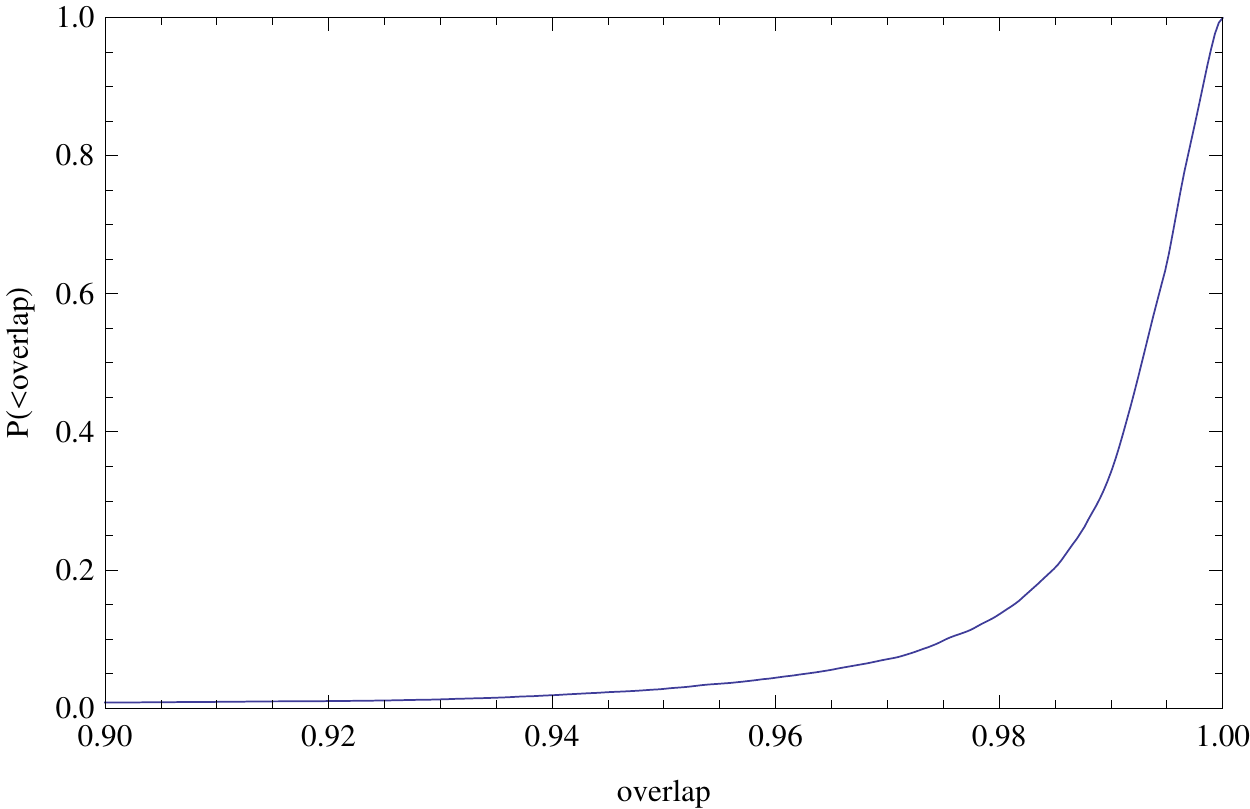}
\caption{\label{fig:Faithful}\textbf{Faithfulness of our model}:  Fraction of  $4\times 10^4$ simulated binaries with overlap
  greater than the specified threshold.  The simulated binaries are each $10+1.4 M_\odot$ BH-NS binaries, with random orientations and random BH
  spins with magnitudes in [0.5,1].  The overlap is calculated as the inner product [Eq. (\ref{eq:overlap}) and below],
  maximized over time and phase, between signals with identical physical parameters, drawn from the
  \texttt{SpinTaylorF2SingleSpin} and \texttt{SpinTaylorT2} models, respectively, using a fiducial advanced LIGO noise curve. The agreement is better that $5\%$
  except for a small number of cases  ($2\%$ with  match less than 0.95) with  significant 
  spin-orbit misalignment   $\hat{L}\cdot\hat{S}\lesssim -0.5$ and unfavorable viewing orientations.
This figure can be reproduced using \texttt{lalsimulation} and the scripts provided on \texttt{arxiv.org}.
 }
\end{figure}

We have implemented our waveform as  \texttt{SpinTaylorF2SingleSpin} in the open-source \texttt{lalsimulation} package
and compared it to the  time-domain
\texttt{SpinTaylorT2} approximant \cite{TaylorT2F2Faithful}.  The noise-weighted inner
product of two waveforms $h_1(t)$ and $h_2(t)$ can be expressed in terms of their fourier transforms $\bar{h}(f)$  as \cite{CutlerFlanagan:1994}
\begin{equation}\label{eq:overlap}
(h_1 | h_2) = 4 \text{Re} \int_{f_l}^{f_h} df \; \frac{\bar h_1(f) \bar h_2^*(f)}{S_n(f)} ~,
\end{equation}
and the overlap is $(h_1 | h_2) / \sqrt{ (h_1 | h_1) (h_2 | h_2)}$. We generate both signals with the exact same mass,
spin, and orientation parameters, then maximize the overlap over only the time and phase.   
Figure \ref{fig:Faithful} shows the overlap distribution for our fiducial scenario: 
$4\times 10^4$ randomly chosen $10+1.4 M_\odot$ BH-NS binaries, each with a random source orientation, random BH spin
direction, and random BH spin magnitude $\chi \in [0.5,1]$. For $S_n(f)$, we use the Advanced LIGO high-power zero-detune noise
spectrum\cite{LIGO-aLIGODesign-Sensitivity,LIGO-ADE-RunPlan} with $f_l = 15\rm Hz$ and terminate the signal at
$f_h=6^{-3/2}/M\pi$, the last stable orbit of a BH with the binary's total mass.  The two
models agree to better than $2\%$ in most  cases, with mismatches of up to $10\%$ occuring only for strong
spin-orbit misalignment ($\hat{L}\cdot \hat{S} =\kappa < -0.9$).  Under these extreme conditions, the spins undergo transitional
precession  \cite{ACST} in the detectors' sensitive band,  breaking the simple precession approximation used here.

\HideSection{Evolution of BH-NS binaries}

\HideSubSection{Notation and Coordinates}

In our notation, the components of the binary have masses $m_1$ and $m_2$, with $m_1$ the larger mass;  we also use  $M
= m_1 + m_2$, $\eta = m_1 m_2 / M^2$.  Only the
larger mass has spin, with a dimensionless spin parameter $\chi_1$ and spin $|{\bf S}_1| = m_1^2 \chi_1$.   [Here and
henceforth we adopt $G=c=1$.]  The orbital
angular velocity is denoted $\omega$ and is related to the gravitational wave frequency by $\omega = \pi f$ and to the
orbital velocity by $v = (M \omega)^{1/3}$. 
We omit trivial dependence on the source sky location.   
To define the orientation of the binary with respect to the solar system barycenter, we use the polar angles
$\theta,\psi$ of the total angular momentum
${\bf J} = {\bf L} + {\bf S}$ relative to the line of sight, where $\psi$ is the  angle of ${\bf J}$ projected into the
plane of the sky and $\theta$ is the angle between ${\bf J}$ and the line of sight;  see  \cite{BLOPrecessionPaper1},
\cite{gwastro-mergers-HeeSuk-FisherMatrixWithAmplitudeCorrections}, 
and references therein.   The ${\bf J}$ direction  is nearly fixed during the inspiral; see \cite{ACST}. 

\HideSubSection{Expressing the signal via a co-rotating frame}

The precessing waveform can be decomposed into a non-precessing waveform acted on by a time-dependent rotation \cite{gwastro-mergers-nr-Alignment-ROS-Methods,gwastro-mergers-nr-Alignment-BoyleHarald-2011,gwastro-mergers-nr-ComovingFrameExpansionSchmidt2010,gwastro-mergers-nr-ComovingFrameExpansion-TransitionalHybrid-Schmidt2012}. The rotation accounts for the precession of the orbital angular momentum of the binary.  The waveform in the source frame is
\begin{eqnarray}
\widetilde{h}(t)_{+} - i \widetilde{h}(t)_{\times} 
&= \displaystyle\sum_{\ell, m'} 
\tilde{h}^{\ell, m'}(t) \mathstrut_{-2}\tilde{ Y}_{\ell, m'}(\widetilde{\theta},\widetilde{\phi}) e^{-i m' \Phi(t)} ~.
\end{eqnarray}
where $\Phi$ is the orbital phase and 
where here and henceforth $\tilde{\cdot}$ denotes the frame co-rotating with the precession of ${\bf L}$.    In this frame, the source amplitudes
nearly obey $\widetilde{h}_{\ell, -m} = (-1)^\ell \widetilde{h}_{\ell, m}^{*}$, to the extent spin-dependent higher
harmonics can be neglected; expressions for $\tilde{h}^{lm}$ exist in the literature \cite{gw-astro-mergers-approximations-SpinningPNHigherHarmonics}.

The source frame and inertial frame are related by a time-dependent rotation,
specified by three Euler angles $(\alpha,\beta,\zeta)$.
[For brevity, we omit $(t)$
  here and henceforth; time dependence is understood for all quantities except $\theta,\phi$ and $\psi$.]  The source frame is aligned with the instantaneous angular momentum $\mathbf{L}$; the
inertial frame is aligned with the total angular momentum $\mathbf{J}$ of the
binary.  In addition to the two Euler angles $(\alpha,\beta)$ that express $\mathbf{L}$ relative to  $\mathbf{J}$, we
apply a third Euler angle $\zeta \equiv - \int \cos \beta (d\alpha/dt) dt$ to minimize superfluous coordinate changes
associated with the orbital plane \cite{gwastro-mergers-nr-Alignment-BoyleHarald-2011}.  Keeping in mind the definition
of $\psi$,  the waveform in an \emph{inertial} frame can be expressed as a weighted sum of the co-rotating-frame amplitudes
$\tilde{h}^{lm}$:
\begin{equation}
h_{+} - i h_{\times} =e^{-2i \psi} \displaystyle\sum_{\ell, m',m} D_{m', m}^{(\ell)}(\alpha, \beta, \zeta) \tilde{h}^{\ell, m}(t) \mathstrut_{-2} Y_{\ell, m'}(\theta,\phi) e^{-i m \Phi}
\end{equation}
where $D^{(\ell)}_{m'm}$ is the usual Wigner rotation matrix representation of SU(2)
\cite{gwastro-mergers-nr-Alignment-ROS-Methods,gwastro-mergers-nr-Alignment-BoyleHarald-2011,gwastro-mergers-nr-ComovingFrameExpansionSchmidt2010}.
Because $h_+-ih_\times$ has spin weight $-2$, this expression is proportional to   $\exp ( - 2 i\psi )$.

For the special case of leading-order quadrupole emission ($\tilde{h}_s^{lm}=0$
unless $l=|m|=2$), the real part of the above expression can be rewritten  as \cite{BLOPrecessionPaper1}
\begin{align}
\label{eq:def:h:zExpansion}
h_+ = \frac{2 M \eta}{D} v^2 \mathrm{Re} \bigg[ z(\alpha-\phi; \theta, \psi, \beta) e^{2 i( \Phi-\zeta)} \bigg] ~
\end{align}
where $z(\alpha)$ is a complex variable that captures the modulation due to precession.   Rather than decompose $z$ into
amplitude-phase modulations  \cite{ACST} or use spherical harmonics  \cite{BCV:PTF}, we expand $z$ as
$z =\sum_m z_m \exp(i m \alpha)$ where
\begin{align}
z_m &= \mathstrut_{-2}Y_{2,m}(\beta, 0) \frac{4\pi}{5}
\left[  e^{-2i\psi} \;  \mathstrut_{-2}Y_{2m}(\theta,0) 
+ e^{2i\psi}  \; \mathstrut_{-2} Y_{2-m}(\theta,0) \right] \; .
\end{align}
The $z_m$ are normalized so $z_m=\delta_{m2}$ when $\beta=\theta=\psi=0$.
In this sum, each  term is proportional to  $e^{i m \alpha}$ and is therefore modulated at  a harmonic of the
precession frequency; multiplying the leading-order $\exp - 2 i \Phi$, each term therefore produces a sideband, offset from
the carrier frequency.
While this expression applies in general, we substantially increase its value by adopting coordinates with a
\emph{hierarchy of timescales}, so  $\Phi$ changes on an orbital timescale; $\alpha$ on a radiation reaction timescale;
and $\beta$ and $v$ on the inspiral timescale.  
This separation of timescales occurs naturally in single-spin BH-NS binaries. 
  To an excellent approximation, a single-spin binary undergoes simple precession  \cite{ACST}, where the total angular momentum
direction is fixed; on short (precession) timescales the orbital angular momentum precesses around $\hat{J}$ at a uniform rate; and
on longer (inspiral) timescales the precession cone opening angle gradually increases.   Adopting coordinates aligned with $J$,
the polar angle $\beta$ is identified as the precession cone opening angle
\cite{BLOPrecessionPaper1} and therefore changes slowly.

\HideSubSection{Orbital evolution}
In the co-rotating frame, orbital evolution can be calculated using the instantaneous binary's binding energy $E(v)$ and
the gravitational wave flux $\mathcal{F}(v)$  \cite{gw-astro-PN-Comparison-AlessandraSathya2009}: 
$
\frac{dt}{dv} = - \frac{dE/dv}{\mathcal{F}} ~.
$
Both $E(v)$ and $\mathcal{F}(v)$ are known as post-Newtonian expansions in the velocity $v$  \cite{gw-astro-PN-Comparison-AlessandraSathya2009,AndySpinModelsDiffer}. These are currently known
to order $v^7$ past the leading order for non-spinning terms and $v^5$ in terms involving spin (note that terms at order
$v^6$ and beyond may also have terms containing powers of $\log v$). 
We expand $dt/dv$ as a power series in $v$, to the same
order as we know the flux and energy, to find $t(v)$ as a closed-form series in $v$. We use the additional relation for the orbital phase
$\frac{d \Phi}{dv} = \frac{dt}{dv} \frac{v^3}{M} $
to obtain $\Phi(v)$ in closed form. 

\HideSubSection{Angular momentum evolution}
For the special case of BH-NS binaries, we can analytically solve the spin-precession equations that determine $\vec{L}$
and hence  $\alpha$, $\beta$, and $\zeta$ as functions of the velocity.  These three angles can be substituted into the
rotation operator to transform the waveform from the precessing source frame into the inertial  frame. 
The leading order (``Newtonian'') expression for the orbital angular momentum is
$
\mathbf{L} = \frac{m_1 m_2}{v} \mathbf{\hat{L}}_N ~,
$
and for the total angular momentum is
$
\mathbf{J} = \mathbf{L} + \mathbf{S}_1 ~.
$
The magnitude of the spin is conserved. In terms of the dimensionless spin, it is $\mathbf{S}_1 = m_1^2 \chi \mathbf{\hat{S}}$. The angle between $\mathbf{L}$ and $\mathbf{S}_1$ does not evolve in the single-spin case, giving the conserved quantity $\kappa = \mathbf{L}_N \cdot \mathbf{S}_1$. We can additionally define two ratios
\begin{align}
\gamma &\equiv \frac{|\mathbf{S}_1|}{|\mathbf{L}|} = \Big(\frac{m_1 \chi}{m_2}\Big) ~ v ~; 
\\
\Gamma_{J} &\equiv | \mathbf{J} | / | \mathbf{L} | = \sqrt{1 + 2 \kappa \gamma + \gamma^2} ~.
\end{align}
The global behavior of $\Gamma_J$ is not well-fit by a single low-order polynomial, as it  approaches  $1$ at $v \rightarrow 0$
and is proportional to $v$ in the limit of large $v$.   As a result, expressions involving $\Gamma_J$ are generally not
well-fit by a standard   PN expansion.

In terms of these quantities, the opening angle  $\beta$ of the cone swept out by $\mathbf{L}$ (the ``precession
cone'') is 
$\cos \beta \equiv \mathbf{\hat L}_N \cdot \mathbf{\hat J} = \frac{1 + \kappa \gamma}{\Gamma_J} ~.
$%
The time evolution of $\mathbf{\hat L}_N$ is given by
$
\frac{d \mathbf{\hat L}_N}{d t} = \Omega_p ~ \mathbf{\hat J} \times \mathbf{\hat L}_N ~,
$
which causes $\mathbf{\hat L}_N$ to rotate around $\mathbf{J}$ with an angular rate of
\begin{equation}
\Omega_p = \eta \left( 2 + \frac{3 m_2}{2 m_1} \right) v^5 ~ \Gamma_J ~.
\end{equation}
The definitions of $\alpha$ and $\zeta$ are then
\begin{eqnarray}
\dot \alpha &=& \Omega_p  \; \quad
\dot \zeta = \dot \alpha \cos \beta = \Omega_p \cos \beta~.
\end{eqnarray}
Substituting the definitions used above leads to 
\begin{eqnarray}
\alpha(v) &=& \eta \left( 2 + \frac{3 m_2}{2 m_1} \right) \int v^5 ~ \Gamma_J \left( \frac{dt}{dv} \right) \mathrm{d}v  \\
\zeta(v) &=& \eta \left( 2 + \frac{3 m_2}{2 m_1} \right) \int v^5 \big( 1 + \kappa \gamma \big) \left( \frac{dt}{dv} \right) \mathrm{d}v  ~.
\end{eqnarray}
We use the TaylorT2 \cite{gw-astro-PN-Comparison-AlessandraSathya2009} expression for $dt/dv$ as a power series in $v$ to express $\alpha$ and
$\zeta$ as integrals in $v$ rather than in $t$.    
Despite the non-polynomial behavior in $\Gamma_J$, both integrals can be evaluated term-by-term to produce closed-form
expressions for $\alpha(v)$ and $\zeta(v)$ at least to 3 PN order, where terms of the form $v^n \log v$ appear.   
These closed-form expressions are provided as supplementary online material, and in the \texttt{lalsimulation} code.
Since $\alpha,\zeta$ and the orbital phase  %
influence each terms' phase in a similar way but  $\alpha,\zeta$ depend on integrals over $dt/dv$
multiplied by at least $v^2$, at least two fewer terms in a post-Newtonian series expansion for $dt/dv$ are needed to
reproduce the precessional dynamics of $\alpha,\zeta$ at the accuracy needed.

\HideSubSection{Stationary phase approximation}

For nonprecessing signals, a commonly-used approximation to a Fourier transform is provided by the stationary-phase
approximation, which for a real-valued signal  $\text{Re}A \exp(i \bar{\Phi}(t))$  has the form  $A(v(f))/2\sqrt{i d^2 \bar{\Phi}/dt^2} \exp i \psi(f)$, where 
where $\psi(f) = 2 \pi f t(v) -  \bar{\Phi}(v)$; see, e.g., Eq. (2.18) in \cite{CutlerFlanagan:1994}.     
Using this approximation, the  Fourier transform of our modulated waveform [Eq. (\ref{eq:def:h:zExpansion})] can be efficiently
computed term-by-term, in general using the phase $\bar{\Phi}_m= 2(\Phi-\zeta)+  m\alpha$.  
At the level of approximation used here, the relation $t(f)$ and hence $\Psi(f)\equiv 2\pi f t(f) - 2\Phi(t(f))$ is
independent of $\alpha$; the factor $\exp(i(-2\zeta+m\alpha))$ can be viewed as a slowly-varying term, like the prefactor $A$.  
  Following custom in gravitational wave data analysis, we approximate  the
 stationary phase amplitude factor $\propto 1/\sqrt{d^2 \bar{\Phi}/dt^2} \simeq (d^2\Phi/dt^2)^{-1/2} $ by its leading order term,  proportional to $f^{-7/6}$.  We therefore find
\begin{align}
\bar{h}_+(f)
&\simeq \frac{2M\eta}{D \sqrt{d^2\Phi/dt^2}}v^2 \sum_m z_m e^{i(\Psi-2\zeta) +im\alpha} \\
&\simeq \frac{2\pi {\cal M}_c^2}{D} \sqrt{\frac{5}{96\pi}} (\pi {\cal M} f)^{-7/6} \sum_m z_m  e^{i(\Psi-2\zeta) +im\alpha} 
\end{align}
where the expressions for $\alpha(v(f))$ and $\zeta(v(f))$ follow by explicit substitution.
The complete expression for $\Psi(f)$ can be found in Eq. (3.18) of \cite{gw-astro-PN-Comparison-AlessandraSathya2009} or Appendix B3 of \cite{AndySpinModelsDiffer}.

\HideSection{Comparisons and Conclusions}

Our fast, faithful frequency-domain waveform provides an efficient,  analytically-tractable representation of gravitational waves from
generic precessing BH-NS binaries.    
Our method generalizes naturally to include higher harmonics. 
As noted above, this method allows for fast, accurate analytic Fisher matrices, to estimate the performance of
parameter estimation.  
More broadly, overlaps between generic single-spin sources can both be calculated and understood analytically.  As a concrete example, 
\citet{BLOPrecessionPaper1} evaluated the overlap between a generic precessing signal and a nonprecessing search
template.   In our representation, each sideband has a unique time-frequency trajectory, offset from the ``chirp'' of the orbital frequency
versus time.   Hence, the best fit between a nonprecessing model will occur when the
 nonprecessing model's time-frequency path lies on \emph{one} of the sidebands; a nonprecessing search misses all power, except
that associated with the optimal sideband; and the best-fitting match will be proportional to $|z_m|$, reproducing their
general result.
Finally, our model suggests several strategies for a viable search for precessing single-spin signals. 
First and foremost, our model allows us to identify precisely which masses,
spins, and viewing orientations would benefit from a multi-modal search.  
Second, the functional form of $z_m$ ensures that usually  two or fewer  $m$ will contribute significantly to the
signal. Each of the harmonics is essentially a nonprecessing waveform. 
A promising search strategy would be to perform a nonprecessing search, then recombine the power from two (or more)
suitable spots.
Third, our expression simplifies the form of overlaps between the $l,m$ modes of precessing signals.  Combined with
a massive reduction in computational cost, our model may allow effective searches for generic precessing systems,
using physical templates and maximizing over source orientation
\cite{BCV:PTF}.

 Figure \ref{fig:Faithful} suggests our approximation is effective.  A subequent publication will expand on the
 material provided here and online, providing a detailed
analysis and error estimates, identifing limits of our approximation and areas for improvement (e.g., better amplitudes
via expanding $d^2\Phi/dt^2$ beyond leading order), exploring its
utility for data analysis and parameter estimation, and discussing the accuracy of its assumptions (e.g., our model for $\alpha(t)$ and $\alpha(f)$;
simple precession;  et cetera).  

\noindent \textbf{Acknowledgements}
ROS is supported by NSF award PHY-0970074. We thank Will Farr and Thomas Dent for helpful discussions and the referees
for careful reading and helpful feedback.

\bibliography{paperexport}

\end{document}